\documentclass{aa501} 
\begin{document}

   \title{The truncation of stellar discs: the magnetic hypothesis}

   \author{E. Battaner \inst{1}, E. Florido\inst{1}  \and  J. Jimenez-Vicente\inst{2}}

   \offprints{E. Battaner}

   \institute{Departamento de F\'{\i}sica Te\'orica y del
              Cosmos, Universidad de Granada, Spain\\
              email: battaner@ugr.es  estrella@ugr.es
         \and
              Kapteyn Institute, Landleven 12, 9747 AD Groningen, The
Netherlands\\
email: jjimenez@astro.rug.nl}
   \date{}
\authorrunning{E. Battaner et al.}

\abstract{We propose a hypothesis of the truncation of stellar discs based on
the magnetic model of the rotation curve of spiral galaxies. Once the
disc had formed and acquired its present structure, approximately, three
balanced forces were acting on the initial gas: gravity and
magnetic forces, inwards, and centrifugal force. When stars
are formed from this gas, the magnetic force is suddenly
suppressed. Gravitation alone cannot retain the newly-formed stars and
at birth places beyond a certain galactocentric radius they
escape to intergalactic space. This radius is the so-called ``truncation
radius'', which is predicted to be at about 4-5 disc radial scale
lengths, in promising agreement with observations.
      }
\maketitle

\section{Introduction}

At large radii,
the surface brightness of a disc no longer follows the standard
exponential profile, but vanishes at the truncation radius, $R_\mathrm{t}$. This
truncation does not take place for the gas density, however, which
continues to decrease exponentially up to much larger radii than $R_t$. 

There is a short but interesting list of works reporting optical
observations of this phenomenon (e.g. van der Kruit 1979; van der
Kruit \& Searle 1981a,b, 1982; Barteldrees \& Dettmar 1994; Pohlen et
al. 2000; Pohlen 2001; de Grijs et al. 2001, hereafter
called dGKW01). Truncations have also been observed in the near
infrared (Florido et al. 2001).

Truncations were originally assumed to be sharp, but recent
observations (Pohlen et al. 2000; dGKW01) have shown most of them to
be fairly shallow.

As most (if not all) discs have stellar truncations, theoretical work
is needed to explain this common feature. However, a recent review by
van der Kruit (2001) noted the scarcity of theoretical scenarios to
explain the basic facts. The aim of this paper is to propose a simple
hypothesis of magnetic origin for the establishment of truncations in
stellar discs. 

\section{Some comments on other hypotheses}

In this section we will comment on other previous hypotheses, their
difficulties and their interest, but it is to be emphasized that our
main purpose is neither a review of these theories, nor a list of
objections about them, but simply to present a novel hypothesis in
order to enrich a future discussion to understand stellar truncations. We
follow the scheme of theoretical scenarios as in dGKW01. These authors
included a comprehensive review in the paper
presenting their own observations, and commented on three alternative
hypotheses. The reader is addressed to dGKW01 for more details and for a
complete list of papers contributing to the development of these hypotheses.

a) {\bf Subcritical gas densities}. Kennicut (1989), among others, holds
that star formation ceases beyond $R_\mathrm{t}$ due to the existence of a
critical gas density, below which star formation does not take place. The
existence of this critical density may be due to various dynamic
mechanisms.

b) {\bf Slow disc formation}.
Following the early models of Larson (1976), Gunn (1982) and others,
discs are still being formed. Intergalactic matter is still accreting
in such a way that the disc radius is an increasing function of
time. The truncation radius is thus the radius where the disc
formation time equals the present age of the galaxy.

c) {\bf Tides}. As shown by Noguchi and Ishibashi (1986) tidal
interactions can also produce the truncation of stellar discs.

dGKW01 propose that a combination of scenarios a) and b) best matches
the observations, i.e. a subcritical gas density in a slowly growing
disc.

Before proposing our own hypothesis, let us comment on the above three.
As observational information is still insufficient no theory can be
firmly rejected which, on the other hand is far from our objectives. 
However, let us note some problems.

As remarked by Sasaki (1987) and dGKW01, tides may explain some
truncations in interacting galaxies, but not all truncations.

The existence of a subcritical gas density is apparently belied
by two observational facts.

In the first place, some galaxies present a sudden step in the rotation
curve. Clear examples of this are NGC891, NGC4013 and NGC5907. The rotation
velocity drops by about 20 km/s at the truncation radius (Bottema et al.
 1987; Bottema 1995, 1996). The implications
of a truncated disc on the rotation curve have been considered by
Casertano (1983), Hunter et al. (1984), Bahcall (1983) and
others. Van der Kruit (2002) and dGKW01 have emphasized the importance
of this fact, although it has yet to be statistically confirmed.

In the absence of internal radial motions, star formation implies an
increase in the stellar density and an equal decrease in the gas
density. However, the total mass density must remain unchanged. Only steps
in the total mass density can produce steps in the rotation
curve. Therefore, a sudden decrease in the star formation rate cannot
explain the sudden step in the rotation curve and, therefore, does
not explain truncations. It is true, however, that this observational
fact needs further studies to be confirmed. Therefore this potential
objection requires more work to be considered.

Secondly, there is in fact star formation beyond $R_t$. Important
stellar formation is observed in the outer disc of the Milky Way. The
truncation radius in our Galaxy is located at about 15 kpc (Habing
1988; Robin et al. 1992; Ruphy et al. 1996; Freudenreich 1998; Porcel
et al. 1997). At radii much larger than this, HII regions, IRAS
sources, H$_2$O masers and other objects characterizing star formation
are observed (e.g. Mead et al. 1987; Mead et al. 1990; Brand \&
Wouterloot 1991, 1994, 1995; Rudolph et al. 1996; Williams \& McKee
1997; May et al. 1997; Kobayashi \& Tokunaga 2000). The amount of star
formation per unit mass of H$_2$ is similar (Wouterloot et al. 1988;
Ferguson et al. 1998), as is the efficiency of star formation
(Santos et al. 2000).

Summarizing these observations with respect to our discussion, it could
be said that H$_2$ could decrease exponentially with radius; 
molecular clouds for $R>R_\mathrm{t}$ would be similar; just the number of 
molecular clouds
should decrease exponentially. For star formation, it is the density 
inside the clouds that is
important, not the number density of clouds. Therefore, a subcritical
gas density could be not confirmed in our own Galaxy.

These arguments are perhaps too naive or based on insufficiently
established observational facts. For example, Martin \& Kennicutt
(2001) have shown that star formation thresholds do exist in nearly
all of the 32 galaxies they studied, even if random star
formation can still occur in isolated places, such as in molecular
clouds at large galactocentric radii. Therefore the threshold
hypothesis cannot be disregarded at all.

The hypothesis of slow disc formation is very attractive: stars are
born at $R > R_\mathrm{t}$, but we do not see (old) stars because this birth
is a very recent phenomenon. The only objection could be the relatively
poor theoretical support. Models by Larson (1976) and Gunn (1982)
were very promising years ago, but present hierarchical cold dark matter
models present a different scenario. Ellipticals are born from the merging
of spirals. These models provide very good results for a large variety
of observational facts but demand substantial revision to form discs
(Navarro \& Steinmetz 2000). The origin of the angular momentum of spiral
discs is not fully understood (Frenk et al. 1997). Gaseous discs in
simulations have much smaller radii than those observed (Navarro et
al. 1995), a problem also present in Larson's models.

The time evolution of disc sizes is difficult to
observe. Bottema (1995) found a relation between truncations and
warps. Many large warps have been found in the HDF (Reshetnikov et
al. 2002) but a direct study of truncations in the HDF has not been
addressed. For z=0 galaxies, Sasaki (1987) pointed out that the mean age
of stars in NGC5907 decreases with increasing galactocentric
distance. This is what would be expected from a slow growth of the
disc, as stated by dGKW01.

The basic question about the truncation problem is then: if stars form beyond 
$R_\mathrm{t}$, why do we only see young stars there? There are two possible
answers: a) The disc for $R>R_\mathrm{t}$ has formed recently, i.e. the
hypothesis of slow disc formation, and b) Stars, once formed, have
gone away. Our model follows this second possibility. It may well be
that both (or more) processes are at work, without a single dominant
mechanism.

\section{The magnetic hypothesis}

We adopt the so-called magnetic model of the rotation curve (Nelson
1988; Battaner et al. 1992; Battaner \& Florido 1995, 2000; Battaner
et al. 1999). In this model, magnetic fields explain the flat rotation
curve, and the model is therefore in disagreement with the standard
interpretation based on the existence of dark matter halos.

These models only provide a first order quantitative output for the
complex problem of the rotation of spirals, though dark matter models
do not do much better. Even so, the magnetic scenario remains an interesting
alternative because of its direct connection
with stellar truncation.

We consider the following simplified scenario: the equilibrium of the
initial gas in the radial direction arises from the balance of two
centripetal forces, gravitational and magnetic, and the centrifugal
force, $\theta^2/R$. (In fact, the magnetic force can be directed
inwards when the magnetic tension is higher than the magnetic pressure
force, as shown in the above-mentioned model). When gas is converted
into stars, the magnetic force suddenly disappears from the stellar
system. Therefore, stars in the outer region suddenly have a velocity
higher than the escape velocity of the gravitational potential. This
escape produces truncation and takes place for stars beyond
a certain radius, which is the observational truncation radius,
$R_\mathrm{t}$. Gas exists beyond $R_\mathrm{t}$ and, therefore, so do newly formed stars,
which can be observed. This young stellar population, however, is assumed to
be in the process of escaping.

A complete model developing this idea should take into account the
time evolution of the system. The redistributed mass in a certain step
should control the dynamics of the next step, in an iterative
process. Instead, we prefer a time-integrated model, which is more suitable for
exploration and zero-order calculations.

Let us assume that, first, there was only gas, and that the density distribution was
exponential. The rotation velocity was different in three regions: a)
an inner region where $\theta(R)$ was that corresponding to an
exponential disc, $\theta_\mathrm{d}(R)$ (with a maximum at 2.27 $R_\mathrm{d}$, 
$R_\mathrm{d}$ being the radial scale length), b) an intermediate region, and c) an
external region where $\theta$ was a constant $\theta_0$. The magnetic
model is fully assumed, and therefore no dark matter is considered,
i.e., the constancy of $\theta_0$ is the result of magnetic effects.

Finally, there are only stars. The density is different in three
regions: a) an inner region, in which magnetism is insignificant
against gravity and, therefore, has remained unchanged,
i.e. exponential, b) an intermediate region, and c) an outer region,
beyond $R_\mathrm{t}$, in which the stellar system density is zero. The rotation
velocity in the inner region is unchanged, $\theta_\mathrm{d}(R)$.

In most galaxies an extended HI disk is observed out to at least twice the
optical radius. The fact that this HI disc exists must mean that
gravity is non-negligible throughout, even in the presence of magnetic
forces in the gas. We have considered, however, that the gravitational
potential is mainly produced by the inner disc ($R < R_\mathrm{t}$).

Taking the thesis of Begeman (1987) as a classical reference, dark
matter is considered to be unnecessary for $R \leq 3R_\mathrm{d}$,
typically. Replacing the dark matter interpretation by the magnetic
one, we assume that magnetic fields are unimportant for $R \leq
3R_\mathrm{d}$, and therefore, the truncation radius must be larger, $R_\mathrm{t} >
3R_\mathrm{d}$. The density for $R < 3R_\mathrm{d}$ contains 95$\%$ of the mass of an
exponential disc, and therefore the gravitational potential beyond $3R_\mathrm{d}$
cannot change very much. This means that for large radii, the potential would not
differ greatly from the central point mass potential and for radii
slightly higher than $3R_\mathrm{d}$ the potential would not differ very much 
from the exponential disc potential. We will estimate $R_\mathrm{t}$ for both
potentials. In each case the calculation is based on the assumption that the
truncation radius will occur when the rotation velocity of the
initial gas becomes equal to the escape velocity: $\theta_{\mathrm{esc}} =
\theta(R_\mathrm{t})$.

\subsection{Central mass potential}

When the potential $\phi =(GM)/(rR_\mathrm{d})$, and $r=R/R_\mathrm{d}$, i.e. the
radius taking $R_\mathrm{d}$ as unity, and the escape velocity $\theta_{\mathrm{esc}}=
\sqrt{2\phi}$, the truncation radius will occur when the velocity
of the initial gas equals this escape velocity. For $R>3R_\mathrm{d}$ it becomes
reasonable to assume that we are in the region where $\theta(R) =
\theta_0$ is a constant. Therefore we can calculate $R_\mathrm{t}$, when
$\theta_0 = \theta_{\mathrm{esc}} (R_\mathrm{t})$. A first simple formula to calculate
the truncation radius is therefore
\begin{equation}
  r_\mathrm{t} ={{2GM} \over {\theta_0^2 R_\mathrm{d}}}
\end{equation}
($r_\mathrm{t} =R_\mathrm{t}/R_\mathrm{d}$, taking $R_\mathrm{d}$ as the unit). Another way to write this
formula could take into account that
\begin{equation}
  {{\theta_{\mathrm{max}}} \over {\sqrt{GM/R_\mathrm{d}}}} = 0.62
\end{equation}
where $\theta_{\mathrm{max}}$ is the maximum velocity, at $R= 2.27 R_\mathrm{d}$, in the
exponential disc rotation curve (see, for instance, Binney \&
Tremaine 1987). Then
\begin{equation}
  r_\mathrm{t} ={2 \over {0.62^2}} \left({{\theta_{\mathrm{max}}} \over {\theta_0}}
  \right)^2 = 5.2 \left({{\theta_{\mathrm{max}}} \over \theta_0} \right)^2
\end{equation}

Here, $\theta_{\mathrm{max}}$ should be taken as the maximum disc circular
velocity from the standard decomposition of $\theta(R)$ into the
different galactic components. Usually, $\theta_{\mathrm{max}}$ is of the order
of $\theta_0$, or slightly lower than $\theta_0$. For instance, van der
Kruit (2002) finds that the ratio $\theta_{\mathrm{max}}/\theta_0$ is
proportional to the square root of $R_\mathrm{d}/H$ ($H$ being the width of
the disc), proposing a mean value of
0.8-0.9. In the sample in
Begeman's thesis containing 8 galaxies, 4 galaxies are pure disc (bulge
poor) galaxies, and hence especially suitable for comparison with our
calculations. These values
are given in Table I.

   \begin{table*}
      \caption[]{}
         \label{}
      \[
         \begin{array}{|c|c|c|c|c|c|}
            \hline
          Name & \theta_{\mathrm{max}} & \theta_0 & \theta_{\mathrm{max}}/\theta_0 &
      r_\mathrm{t}=5.2(\theta_{\mathrm{max}}/\theta_0)^2 &
      r_\mathrm{t}{\cal B}(r_\mathrm{t})=(\theta_0/\theta_
        {\mathrm{max}})^20.38\\
          & \mathrm{(km/s)} & \mathrm{(km/s)} & &\mathrm{(central\ mass\ point)}&\mathrm{ (exponential\ disc)} \\
           \hline
      \mathrm{NCG} 3198 & 150 & 150 & 1.00 & 5.2 & 5.6\\
      \mathrm{NGC} 2403 & 120 & 140 & 0.86 & 3.8 & 4.2\\
      \mathrm{NGC} 2903 & 210 & 190 & 1.11 & 6.4 & 6.7\\
      \mathrm{NGC} 6503 & 110 & 130 & 0.85 & 3.8 & 4.1\\ \hline
      \end{array} 
      \]
   \end{table*}

NGC 2903 was considered somewhat exceptional by Begeman. We see that truncation
radii between 5.2 and 3.8 $R_\mathrm{d}$ are typical with this approximation.

\subsection{Exponential disc potential}

Now
\begin{equation}
\begin{array}{rl}
  \phi(R) & = -2\pi G \Sigma_0 R {\cal B}(r) =
  - {{GM} \over {2R_\mathrm{d}^2}} R {\cal B}(r) \\
  &  \\
  &  =  -{1 \over 2}{{GM} \over R_\mathrm{d}} r {\cal B}(r) =
  - {1\over 2} {{\theta_{\mathrm{max}}^2} \over {0.62}^2} r {\cal B}(r)
\end{array}
\end{equation}
where $\Sigma_0$ is the disc central surface brightness and
\begin{equation}
  {\cal B}(r) = I_0 (r/2) K_1(r/2) - I_1 (r/2) K_0(r/2)
\end{equation}
where $I_0$, $I_1$, $K_0$ and $K_1$ are Modified Bessel
functions. Again, we calculate $r_\mathrm{t}$ considering $\theta_{\mathrm{esc}}(r_\mathrm{t})=
\theta_0$:
\begin{equation}
  r_\mathrm{t} {\cal B}(r_\mathrm{t}) = \left( {{\theta_0} \over {\theta_{\mathrm{max}}}}\right)^2
  0.38
\end{equation}

In the case of $\theta_{\mathrm{max}} = \theta_0$ we obtain $r_\mathrm{t} =5.6$. In Table
1, we see that with this potential we
obtain slightly larger values in the range $r_\mathrm{t}= 4.1-5.6$ if we still
consider NGC2903 somewhat exceptional.

Giving the uncertainties in our simple model, the results are very
promising. Some values reported in the literature are: van der Kruit
\& Searle (1982), 4.2 $\pm$ 0.5; de Grijs et al. (2001), 4.3, 3.8, 4.5
and 2.4; Pohlen et al. (2000 a,b), 2.9 $\pm$ 0.7; Barteldrees \&
Dettmar (1994), 3.7 $\pm$ 1. Data by
Florido et al. (2001) were obtained in the NIR, minimizing
extinction effects, especially those in the Ks band. By application of
this filter,
they obtained $r_\mathrm{t} = 3.6 \pm 0.8$. 

The agreement between
observations and the values deduced here is therefore very close.

\subsection{The decoupling time}

Throughout this paper we have assumed that the time scale for the
formation of stars out of a gas cloud is short compared to the
galactic dynamic time. In this section we will check the validity
of this assumption.

In the transition from molecular gas to a main sequence star there
must be three different regimes. In the first one, external magnetic
fields are still dynamically important, while in the third one (close
to the main sequence stage) they are completely ignorable. Therefore, the
timescale of the intermediate stage (or the time of decoupling from
external magnetic fields) is a good estimate of how fast magnetic
fields stop being dynamically relevant. If this decoupling time is
large enough, then complicated orbits would develop during this 
intermediate phase which
would affect the degree of sharpness of the truncation.

A 1M$_\odot$ star reaches the main sequence in approximately $10^7$
years. If the star population is dominated by lower mass stars, this time could
be as long as $10^8$ years. However, the
decoupling time must be much shorter. In fact, before nuclear reactions begin,
the proto-star reaches a very high density. In this stage self-gravity 
becomes so high that external magnetic fields cannot modify the orbit
of such a dense system. This high density is reached after the initial
free fall collapse. Therefore, the free fall time can be used to
estimate the decoupling time

\begin{equation}
  \tau_{\mathrm{ff}} = \left( {{3\pi} \over {32 G \rho}} \right)
\end{equation}

Where $\rho$ is the density of the initial molecular cloud. 

This time
is independent of the mass of the born star, the size of the cloud
and even the fragmentation process. For a characteristic cloud
density of $5 \times 10^{-22} \mathrm{g} \mathrm{cm^{-3}}$, the free fall time is $3
\times 10^6$ years.

Desch and Mouschovias (2001) have calculated that the cloud does not
actually suffer a free fall collapse. Instead, the collapse is slowed down
approximately 30$\%$ due to the magnetic pressure. This results 
in a time of about $10^7$ years which is, in fact, an 
upper limit to the decoupling time.

Desch and Mouschovias (2001) have also considered a complementary
problem. Instead of the external problem considered here of when the orbit
of the protostar becomes unaffected by large scale galactic fields,
they have considered when internal magnetic fields are negligible in
the collapse process. This complementary calculation can also be
used here.

They start with a cloud density of $3 \times 10^2 \mathrm{cm^{-3}}$. About 13
Myr later the density has only increased less than two orders of
magnitude. Along these 13 Myr magnetic fields are internally important
and keep values not much higher than interstellar strengths. Suddenly,
the density increases, and after 15 Myr the number density becomes much
larger (with typical values of $10^{12} \mathrm{cm^{-3}}$). The action of
ambipolar diffusion
prevents the increase of magnetic fields, which therefore become
negligible in the collapse dynamics. It is clear that such a dense 
object cannot be
displaced by external magnetic fields. The sudden increase of the
density takes place in a quite short period (which can be taken as
an estimate for the
decoupling time) of $2 \times 10^6$ years.

Moreover, the rotation velocity of the proto-star should also suddenly
increase. Through reconnection of magnetic field lines, the external
and internal magnetic fields should become independent. Interstellar
magnetic field lines are not able to penetrate into the protostar,
and the external magnetic field is therefore
unable to have an influence on the dynamics.

A decoupling time of $2 \times 10^6$ years is very short compared with
typical dynamic times in the galaxy. Therefore, the transition from
magnetically driven clouds to magnetically unaffected proto-stars can
be assumed to be instantaneous.

\section{Conclusions}

We have considered the different scenarios proposed to date to
explain truncations. Models based on the existence of a cut-off
density, below which the star formation rate vanishes, do not
perfectly agree with observations. Tidal effects cannot explain all
truncations. The scenario based on Larson's (1976) model with a
slowly growing disc is, however, an interesting possibility. 
Some additional theoretical work supporting this model is
desirable. Even if it is not
possible at present to reject any previous hypothesis, new ones would
be welcome. We suggest a hypothesis based on magnetic fields.

The magnetic hypothesis of the truncation of stellar discs, 
presented here, provides a simple explanation and predicts values of the
truncation radius that are in very satisfactory agreement with the
observations. This result may also constitute an argument supporting
the magnetic model of rotation curves. The two phenomena could be closely
related. More sophisticated models are a desirable goal but,
at present, exploratory zero-order arguments are preferable. This is an
interesting alternative, especially considering that not many
scenarios have
been advanced in the literature.

In the model presented here, an escape of stars to the
intergalactic medium is predicted. It is then necessary to estimate the amount of
escaped mass along the whole history of the galaxy.

The mass of lost stars is negligible compared with the initial mass of
the galaxy. It can be estimated by
\begin{equation}
  \int_{r_\mathrm{t}}^\infty \rho_0 e^{-R/R_\mathrm{d}} 4 \pi RH dR
\end{equation}
where $\rho_0$ is the initial central density and $H$ the width of the
disc. Therefore, the proportion of lost stars would be
\begin{equation}
  {{\left[ \int_{r_\mathrm{t}}^\infty e^{-r} r\,dr\right]} \over
  {\left[ \int_0^\infty e^{-r} r\,dr\right]} } \approx 0.1
\end{equation}
when adopting $r_\mathrm{t}=4$. Therefore, only 10$\%$ of
the initial galactic mass escapes to the intergalactic space due to
this mechanism, during
the whole life-time of a galaxy. If gravitational potentials in the
outer galaxy, other than that considered here would not be negligible,
this 10$\%$ would be an upper limit.

Considering Eq. (1), some tentative predictions can be made about
the dependences of $r_\mathrm{t}$, which could be tested in future statistical
studies: the  truncation radius should roughly be inversely
proportional to the squared asymptotic rotation velocity $r_t \propto
\theta_0^{-2}$. To find a predictable relation of $r_\mathrm{t}$ on $M$, and
$R_\mathrm{d}$, we should take into account that  $M
\propto R_\mathrm{d}^2$, and therefore $r_\mathrm{t}$ should be proportional either to $R_\mathrm{d}$, or to
the square root of the visible mass.
Similar conclusions are obtainable from Eq. (4).

On qualitative grounds, other statistical properties are to be expected
from this hypothesis. We predict sharper truncations and lower
truncation radii for larger wavelengths. In fact, the old stellar
population has been affected but its escape has been considered here
for 
the whole history of a galaxy. Recent star formation beyond $r_t$,
however, would produce a much smoother decline among the young population. Indications of this
colour dependence of both sharpness and radius are evident in the
comparison of the optical data obtained by Pohlen (2001) and Florido et
al. (2001).

\begin{acknowledgements}
This paper has been supported by the ``Plan Andaluz de Investigaci\'on''
(FQM-108) and by the ``Secretar\'{\i}a de Estado de Pol\'{\i}tica
Cient\'{\i}fica y Tecnol\'ogica'' (AYA2000-1574).
\end{acknowledgements}

\end{document}